# SLDs for Visualizing Multicolor Elevation Contour Lines in Geo-Spatial Web Applications

Kodge B. G.
Department of Computer Science
Swami Vivekanand Mahavidhyalaya, Bidar Road,
Udgir, Dist. Latur (MS), India
kodgebg@hotmail.com

Hiremath P. S.
Department of Computer Science
Gulbarga University, Gulbarga
Karnataka State, India
Hiremathps53@yahoo.com

Abstract— This paper addresses the need for geospatial consumers (either humans or machines) to visualize multicolored elevation contour poly lines with respect their different contour intervals and control the visual portrayal of the data with which they work. The current OpenGIS Web Map Service (WMS) specification supports the ability for an information provider to specify very basic styling options by advertising a preset collection of visual portrayals for each available data set. However, while a WMS currently can provide the user with a choice of style options, the WMS can only tell the user the name of each style. It cannot tell the user what portrayal will look like on the map. More importantly, the user has no way of defining their own styling rules. The ability for a human or machine client to define these rules requires a styling language that the client and server can both understand. Defining this language, called the StyledLayerDescriptor (SLD), is the main focus of this paper, and it can be used to portray the output of Web Map Servers, Web Feature Servers and Web Coverage Servers.

## I. INTRODUCTION

The appearance of a map in terms of 'styled layers' can be considered as a transparent sheet with features symbolized upon it. A map is made up of a number of these styled layers put together in a specified order. The styled layers are said to be Z-ordered. Users can define more complex or simpler maps by adding or removing styled layers.

A styled layer itself represents a particular combination of 'layer' and a 'style' in which that layer can be symbolized. Conceptually, the layer defines a stream of features and the style defines how those features are symbolized. This concept is underlined by the fact that there may be multiple styles in which a layer can be symbolized. In the WMS specification, the request for a map is encoded as an HTTP-GET request and the appearance for a map portrayal is specified by the LAYERS and STYLES parameters.

Consider the following (incomplete) example1 map request (which is split over multiple lines for presentation purposes only):

```
VERSION=1.1.0&
REQUEST=GetMap&
BBOX=0.0,0.0,1.0,1.0&
LAYERS=Rivers,Roads,Houses&
STYLES=CenterLine,CenterLine,Outline
```

Rule(1)

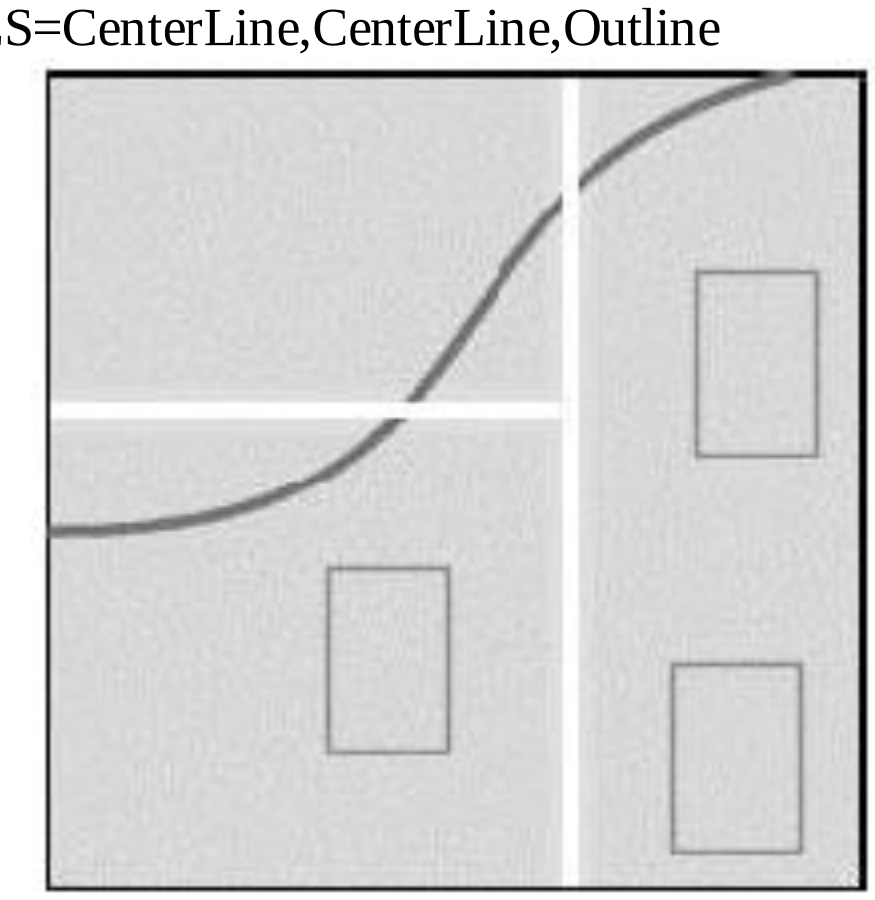

Figure 1. Results in the map portrayal based upon rule (1)

This is to be interpreted as three 'styled layers', namely:

- Houses:Outline
- Roads:CenterLine
- Rivers:CenterLine

The colon notation is introduced only as a convenience to aid discussion. The Rivers:CenterLine styled layer is 'below'

the Roads:CenterLine styled layer, as WMS uses the "painter's model" and plots each successive layer in the LAYER list over top of the previously rendered layers. Consequently, the roads appear to 'cross' the river. It is possible for the same layer to appear more than once, although rarely with the same style. A common 'cartographic trick' to generate what appears to be the boundaries of linear features is to draw them with a thick colored line and then draw them all again with a thinner, lighter line. This is done for the roads in the following example2 (incomplete) map request:

VERSION=1.1.0&
REQUEST=GetMap&
BBOX=0.0,0.0,1.0,1.0&
LAYERS=Roads,Roads,Houses&
STYLES=Casing,CenterLine,Outline Rule(2)

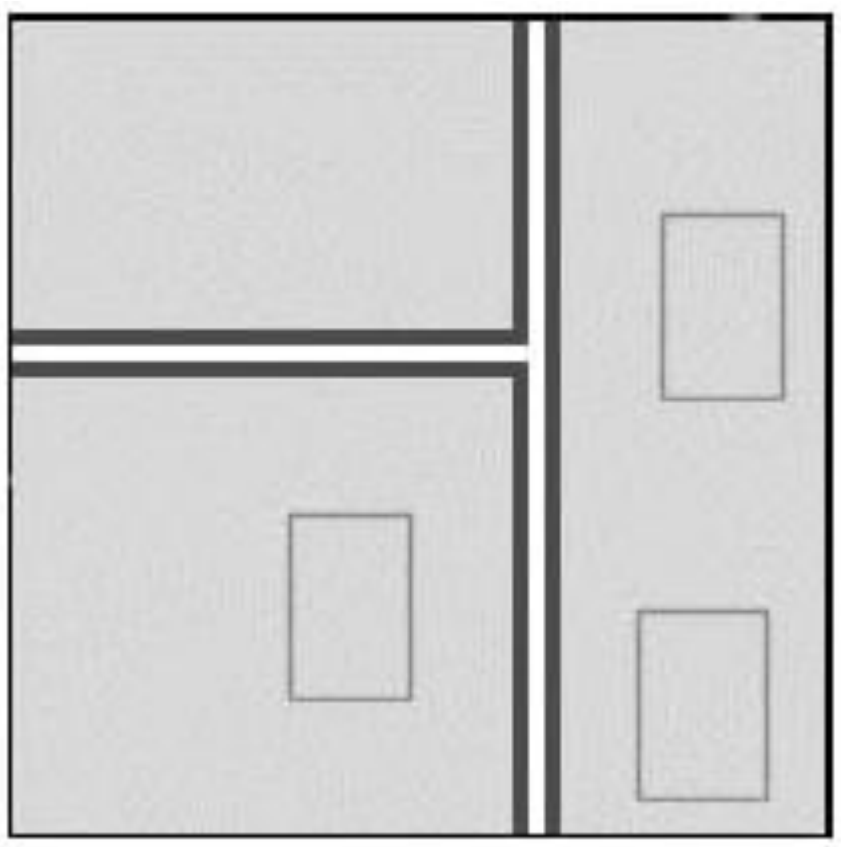

Figure 2. Resulting map portrayal based upon rule (2)

This is to be interpreted as three styled layers, namely:

- Houses:Outline
- Roads:CenterLine
- Roads:Casing

It might be noted that the WMS cannot be interrogated for metadata to indicate which styled layers can be meaningfully combined and how. However, a flexible client would allow an end-user to explore the various possibilities. The WMS 1.1.1 specification deals with styles and layers which are 'known' to the WMS and which are identified by name. For this reason, the rest of this document refers to the layers and styles that have been described above as "named layers" and "named styles". The WMS specification provides only one way to define a styled layer, as a combination of a named layer and a named style.

## II. STYLED LAYER DISCRIPTOR

The appearance of a map in the WMS specification can be defined as a sequence of styled layers. Styling can also be described using a user-defined XML encoding of a map's appearance called a Styled-Layer Descriptor (SLD). The SLD includes a StyledLayerDescriptor XML element that contains a sequence of styled-layer definitions. These styled-layer definitions may use named or user-defined layers and named or user-defined styling. Here is an example simple SLD that corresponds to the first example1 from section I:

```
<StyledLayerDescriptor version="1.0.0">
   <NamedLayer>
      <Name>Rivers</Name>
      <NamedStyle>
         <Name>CenterLine</Name>
      </NamedStyle>
   </NamedLayer>
   <NamedLayer>
      <Name>Roads</Name>
      <NamedStyle>
         <Name>CenterLine</Name>
      </NamedStyle>
   </NamedLayer>
   <NamedLayer>
      <Name>Houses</Name>
      <NamedStyle>
         <Name>Outline</Name>
      </NamedStyle>
   </NamedLayer>
</StyledLayerDescriptor>
```

The NamedLayer and NamedStyle elements correspond to the LAYERS and STYLES of the CGI parameters and the "painter's model" is also used for Z-ordering. An SLD XML document can become much more complex with user-defined styling. The WMS-1.2 styled-layer mechanism is compatible with the SLD-1.0.0 format.

## III. WMS REQUESTS USING AN SLD

Three approaches are defined to allow a client to take advantage of SLD:

- The client interacts with the WMS using HTTP GET but the request can reference a remote SLD.
- The client uses the HTTP GET method but includes the SLD XML document inline with the GET request in an SLD_BODY CGI parameter (with appropriate character encoding).
- The client interacts with the WMS using HTTP POST with the GetMap request encoded in XML and including an embedded SLD.

The third method is technically superior but there has been a great lack of vendor support for the XML-POST GetMap-request method. Use of the second method, which is a compromise between the first and third methods, can encounter problems due to excessively long URLs. It is important to note that in all cases the WMS has no prior knowledge of the SLD contents. There is a wide spectrum of possible clients. Some may allow a user to switch between a number of pre-defined maps, each specified by its own pre-defined SLD. Others may allow a user to interactively define how they wish a map to appear and construct the necessary SLD 'on-the-fly'. All of the approaches described above allow a client application to do this but the first one requires that the client be able to place the SLD document in a Web location accessible to the WMS.

When an SLD is used as a style library, the STYLES CGI parameter is interpreted in the usual way in the GetMap request, except that the handling of the style names is organized so that the styles defined in the SLD take precedence over the named styles stored within the map server. The user-defined SLD styles can be given names and they can be marked as being the default style for a layer. To be more specific, if a style named “CenterLine” is referenced for a layer and a style with that name is defined for the corresponding layer in the SLD, then the SLD style definition is used. Otherwise, the standard named-style mechanism built into the map server is used. If the use of a default style is specified and a style is marked as being the default for the corresponding layer in the SLD, then the default style from the SLD is used; otherwise, the standard default style in the map server is used.

If a WMS is to symbolize features using a user-defined symbolization, it is necessary to identify the source of the feature data. This specification is designed to permit a wide variety of implementations of WMS that support user-defined symbolization. For example a WMS might symbolize feature or coverage data stored in a remote Web Feature Server (WFS) or Web Coverage Server (WCS), or it might only be able to symbolize data from a specific default feature/coverage store. In support of this, optional parameters called REMOTE_OWS_TYPE and REMOTE_OWS_URL are introduced for HTTP-GET GetMap requests that can be used to direct the WMS to a remote WFS, WCS, or other OWS service as the ‘default’ source for feature/coverage data. The presently allowed values for the REMOTE_OWS_TYPE parameter are “WFS” and “WCS”, though more may be allowed in the future. The REMOTE_OWS_URL parameter gives the base URL of the service to use.

## IV. VISUALIZING MULTICOLOR CONTOUR LINES

Contouring is the most common method for terrain mapping. Contour line connects points of equal elevation, the contour interval represents the vertical distance between contour lines, and the base contour is the contour from which contouring starts. Contour lines are lines drawn on a map connecting points of equal elevation. The contour line represented by the shoreline separates areas that have elevations above sea level from those that have elevations below sea level. We refer to contour lines in terms of their elevation above or below sea level. In this example, the shoreline would be the zero contour line ( it could be 0 ft., 0 m, or something else depending on the units we were using for elevation). Contour lines are useful because they allow us to show the shape of the land surface (topography) on a map. Suppose a DEM has elevation readings from 362 to 750 meters. If the base contour is set to 400 and the contour interval at 100, then contouring would create the contour lines of 400, 500, 600 and so on. Contour lines can be drawn for any elevation, but to simplify things only lines for certain elevations are drawn on a topographic map[1]. These elevations are chosen to be evenly spaced vertically. This vertical spacing is referred to as the contour interval. For example if the maps use a 10 ft contour interval, each contour lines are a multiple of 10 ft.( i.e. 0, 10, 20, 30, etc). Other common intervals seen on topographic maps are 20 ft (0, 20, 40, 60, etc), 40 ft (0, 40, 80, 120, etc), 80 ft (0, 80, 160, 220, etc), and 100ft (0, 100, 200, 300, etc). The contour interval chosen for a map depends on the topography in the mapped area. In areas with high relief, the contour interval is usually larger to prevent the map from having too many contour lines, which would make the map difficult to read.

The contour interval is constant for each map. It will be noted on the margin of the map. One can also determine the contour interval by looking at how many contour lines are between labeled contours.

Unlike the simple topographic map, the real topographic maps have many contour lines. It is not possible to label the elevation of each contour line. To make the map easier to read, every fifth contour line vertically is an index contour. Index contours are shown by darker brown lines on the map. These are the contour lines that are usually labeled [5].

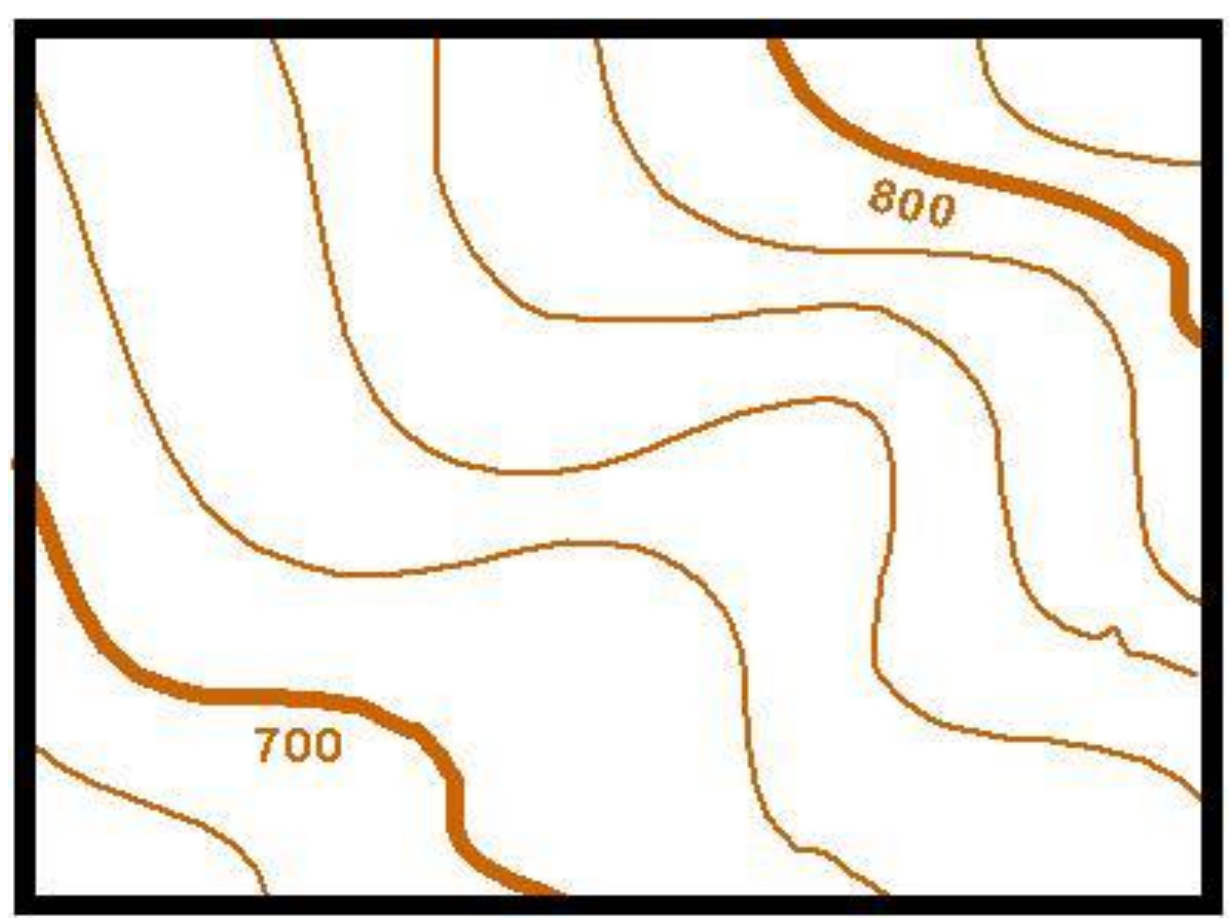

Figure 3. Section of a topographic map.

The example Fig.3 illustrates a section of a topographic map. The brown lines are the contour lines. The thin lines are the normal contours, while the thick brown lines are the index contours. The elevations are only marked on the thick lines

The contour lines of a projected area of having 450 to 700 meters of elevation with 25 meters contour interval are generated. The generated contour lines are next stored in a PostGRESQL spatial relational database with respect to their attributes namely, Geo_ID, Contour, Elevation, Layer.

All the contour lines from the stored spatial data base are imported to GEOSERVER and prepared a SLD in XML to visualize the contour lines in multicolor scale according to the elevation values of each and every contour lines. The Spatial data table of contour lines and a program written in XML to create and visualize a user defined SLD as follows.

TABLE I. ATTRIBUTES OF ELEVATION CONTOUR LINES

| **Geo_ID** | **ELEVATION** | **Contour Line** | **Projected Layer** |
|---|---|---|---|
| C_001 | 450 | POLY_LINE | Binary_Data |
| C_002 | 475 | POLY_LINE | Binary_Data |
| C_003 | 500 | POLY_LINE | Binary_Data |
| . | . | . | . |
| . | . | . | . |
| . | . | . | . |
| C_010 | 675 | POLY_LINE | Binary_Data |
| C_011 | 700 | POLY_LINE | Binary_Data |

Page No. Coding

```
1   <?xml version="1.0" encoding="ISO-8859-1"?>
2   <StyledLayerDescriptor version="1.0.0"
3       xsi:schemaLocation="http://www.opengis.net/sld StyledLayerDescriptor.xsd'
4       xmlns="http://www.opengis.net/sld"
5       xmlns:ogc="http://www.opengis.net/ogc"
6       xmlns:xlink="http://www.w3.org/1999/xlink"
7       xmlns:xsi="http://www.w3.org/2001/XMLSchema-instance">
8     <NamedLayer>
9       <Name>Attribute-based line</Name>
10      <UserStyle>
11        <Title>SLD Cook Book: Attribute-based line</Title>
12        <FeatureTypeStyle>
13          <Rule>
14            <Name>450</Name>
15            <ogc:Filter>
16              <ogc:PropertyIsEqualTo>
17                <ogc:PropertyName>ELEVATION</ogc:PropertyName>
18                <ogc:Literal>450</ogc:Literal>
19              </ogc:PropertyIsEqualTo>
20            </ogc:Filter>
21            <LineSymbolizer>
22              <Stroke>
23                <CssParameter name="stroke">#0000FF</CssParameter>
24                <CssParameter name="stroke-width">1</CssParameter>
25              </Stroke>
26            </LineSymbolizer>
27
28            <TextSymbolizer>
29              <Label>
30                <ogc:PropertyName>ELEVATION</ogc:PropertyName>
31              </Label>
32              <Fill>
33                <CssParameter name="fill">#000000</CssParameter>
34              </Fill>
35              <Font>
36                <CssParameter name="font-family">Arial</CssParameter>
37                <CssParameter name="font-size">8</CssParameter>
38                <CssParameter name="font-style">normal</CssParameter>
39                <CssParameter name="font-weight">bold</CssParameter>
40              </Font>
41              <VendorOption name="followLine">true</VendorOption>
42              <VendorOption name="maxAngleDelta">90</VendorOption>
43              <VendorOption name="maxDisplacement">400</VendorOption>
44              <VendorOption name="repeat">150</VendorOption>
45              <LabelPlacement>
46                <LinePlacement />
47              </LabelPlacement>
48            </TextSymbolizer>
49
50          </Rule>
51        </FeatureTypeStyle>
52
                 •
                 •
                 •
                 •
                 •
466
467
468     </UserStyle>
469   </NamedLayer>
470 </StyledLayerDescriptor>
```

Figure 4. XML code to visualize multicolor contour lines

The figure 4 shows the XML code to create and visualize the contour lines of different elevation data. In this code the line no 17 extract the features of ELEVATION attributes. The values of a particular elevation contour will be compared in line no 18 and applied the respective color to it as mentioned in line no. 23 in hexadecimal color value blue (#0000FF). The size of the displaying contour line will be assigned by line no. 24, i.e the width (thickness) of line.

Apart from the visualization of only poly lines will not be sufficient output for us, the associated contour lines values must be display with their contour lines. Displaying contour values with contour polylines in different angles are mentioned in line no. 41, 42, 43 and 44. and the style of font is mentioned in line no. 36, 37,38 and 39.

All the above discussed code belongs to visualizing a contour poly lines having elevation 450 meters. The same process will apply for all the elevation values of their respective contour poly lines with different required colors and values.

## V. RESULT

After validating the above discussed XML code, we can visualize multicolored set of contour lines with different elevation values (labels) as shown in below figure 5.

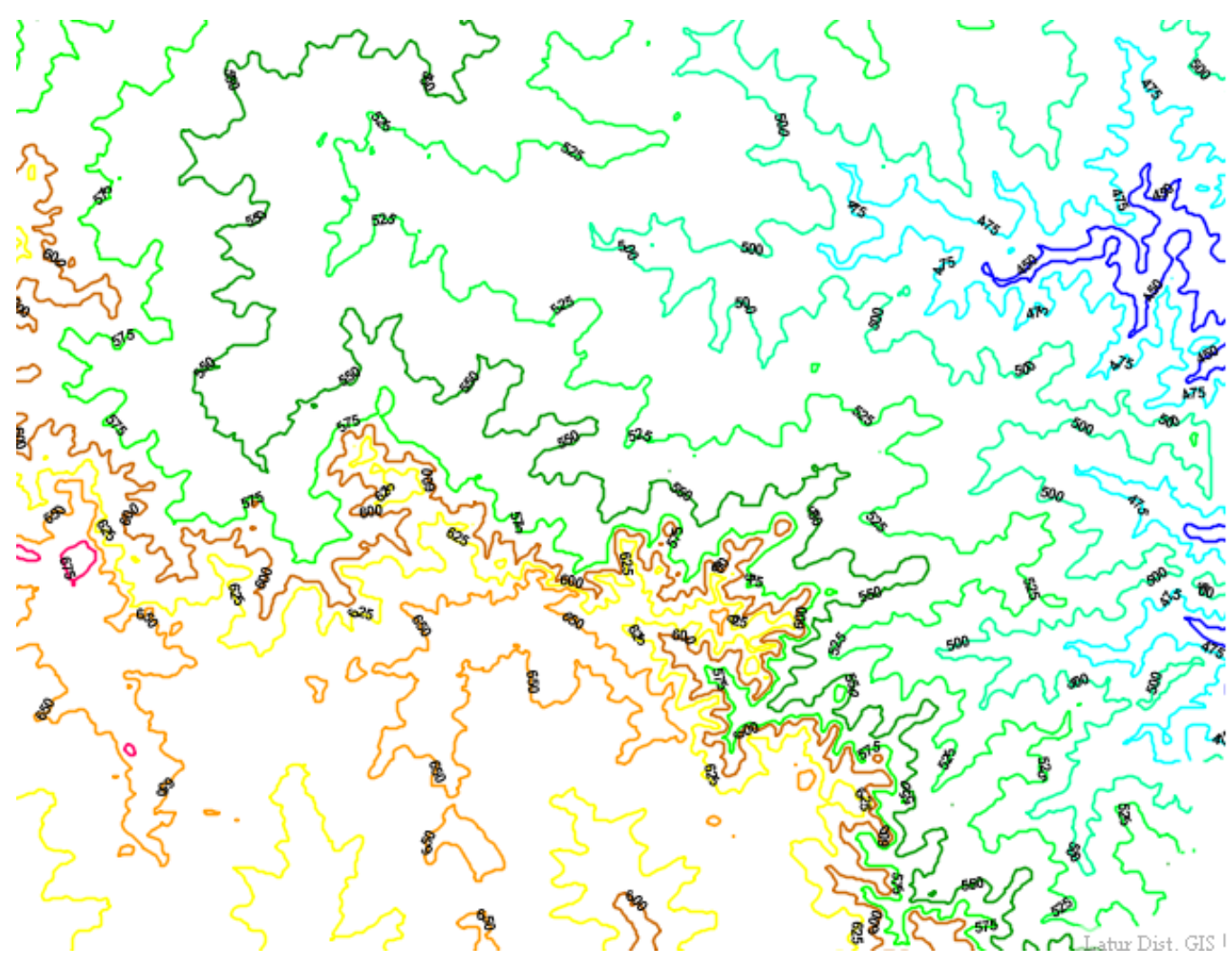

Figure 5. Multicolored Contour lines on web

## VI. ACKNOWLEDGMENT

The authors are indebted to Open Geo-Spatial Consortium for providing such an extensible platform to create user defined web based spatial information system.

## VII. Conclusion

The thematic maps of any projected area consists contour lines to represent the elevation of that particular area. Now a day the geographical information on internet is become very rich and dynamic. So this paper deals with the visualization of elevation contours and their associated values in web based geospatial information system using styled layer descriptors for online spatial information system in web based applications.

## Authors Profile

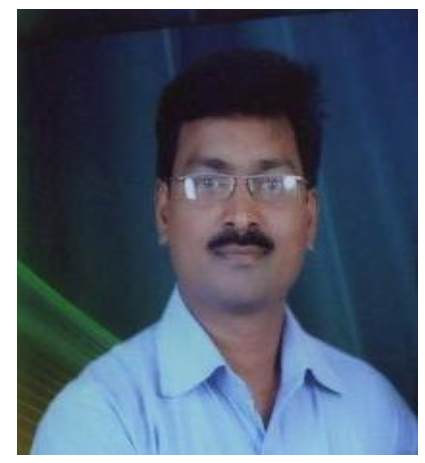

Mr. **Kodge B. G.** is Ph.D. research scholar in Computer Science of Swami Vivekanand College, Udgir Dist. Latur (MH) INDIA. I obtained MCM (Master in Computer Management) in 2004, M. Phil. in Computer Science in 2007 and submitted Ph.D. thesis in computer science in 2011. My research areas of interests are GIS and Remote Sensing, Digital Image Processing and Pattern Recognition, Data mining and data warehousing. I have published more than 28 research papers in reputed national, international journals and proceedings conferences. I have received two research projects from UGC and DST of India. Tel. +919923229672

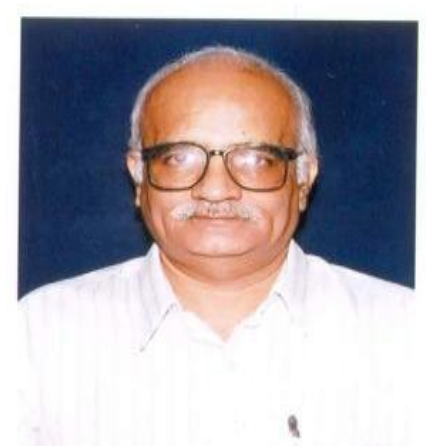

Dr. **Hiremath P. S.** is a Professor and Chairman, Department of P. G. Studies and Research in Computer Science, Gulbarga University, Gulbarga-585106 INDIA, He has obtained M.Sc. degree in 1973 and Ph.D. degree in 1978 in Applied Mathematics from Karnataka University, Dharwad. He had been in the Faculty of Mathematics and Computer Science of Various Institutions in India, namely, National Institute of Technology, Surathkal (1977-79), Coimbatore Institute of Technology, Coimbatore(1979-80), National Institute of Technology, Tiruchirapalli (1980-86), Karnatak University, Dharwad (1986-1993) and has been presently working as Professor of Computer Science in Gulbarga University, Gulbarga (1993 onwards). His research areas of interest are Computational Fluid Dynamics, Optimization Techniques, Image Processing and Pattern Recognition. He has published 152 research papers in peer reviewed International Journals and proceedings of conferences. Tel (off): +91 8472 263293, Fax: +91 8472 245927.